\magnification=1200
\def\setup{\count90=0 \count80=0 \count92=0 \count93=0 \count85=0
\countdef\refno=80 \countdef\secno=85 \countdef\equno=90 \countdef\equnoA=92 \countdef\equnoB=93}
\def\autoref{ {\global\advance\refno by 1} \kern -5pt [\the\refno]\kern 2pt}  
\def\autorefb{ {\global\advance\refno by 1} \kern -9pt \the\refno\kern 1pt}  
\def\autorefs{ {\global\advance\refno by 1} \kern -6.2pt}  
\def\autoeq{ {\global\advance\count90 by 1} \eqno(\the\secno.\the\equno) } 
\def\autoeql{ {\global\advance\count90 by 1} & (\the\secno.\the\equno) }  
\def\autosec{ {\global\advance\secno by 1} \kern -8pt \the\secno\  }
\def\autoeqA{ {\global\advance\count92 by 1} \eqno(A.\the\equnoA) }
\def\autoeqB{ {\global\advance\count93 by 1} \eqno(B.\the\equnoB) }
\def\CPN{ {{\bf CP}^{N-1} } }
\def\CPNF{ {{\bf CP}^{N-1}_F } }
\font\small=cmr10 scaled 750
\font\smallit=cmti10 scaled 750
\def\A{ \hat{\cal A} }
\def\G{ {\cal G} }
\def\H{ {\cal H} }
\def\P{ {\cal P} }
\def\T{ {\cal T} }
\def\M{ {\cal M} }
\def\N{ {\cal N} }

\def\X{ {\bf X} }
\def\Y{ {\bf Y} }
\def\Z{ {\bf Z} }
\def\hA{ {\hat A} }
\def\hB{ {\hat B} }

\def\Tr{ \hbox{Tr} }

\overfullrule=0pt
\setup
\centerline{  }
\line{\hfill \hbox{hep-th/0107099}}
\line{\hfill \hbox{SU-4252-740}}
\line{\hfill \hbox{DIAS-STP-01-11}}
\line{\hfill \hbox{Revision no. 2} }
\line{\hfill \hbox{30th October 2001}}
\vskip 1cm
\centerline{\bf  Fuzzy Complex Projective Spaces and their Star-Products} 
\vskip 1.2cm
\centerline{A.P.~Balachandran,$^{a}$\footnote{}{{\hskip -7pt
\small e-mails: bal@phy.syr.edu, bdolan@thphys.may.ie,
joohan@kerr.uos.ac.kr,  xavier@fis.cinvestav.mx, denjoe@fis.cinvestav.mx}}
Brian~P.~Dolan,$^{b,}$\footnote{$^{c}$}{\small On leave of absence from:
{\smallit Department of Mathematical Physics, 
NUI Maynooth, Maynooth, Ireland}
}
J.~Lee,{$^{d}$} X.~Martin{$^{b}$}
and Denjoe~O'Connor$^{b}$}
\vskip .5cm
\leftline{$^a$ \it Physics Department, Syracuse University, Syracuse, NY 13244-1130, USA}
\leftline{$^b$ \it Depto de F{\rm\'\i}sica, Cinvestav, Apartado Postal 70-543, 
M\'exico D.F. 0730, M\'exico}
\leftline{$^d$ \it Physics Department, University of Seoul, Seoul 130-743, Korea}
\vskip 1.5cm
\centerline{\bf ABSTRACT}
\bigskip
\noindent
We derive an explicit expression for an associative $*$-product on the fuzzy complex
projective
space, ${\bf CP}^{N-1}_F$.  This generalises previous results for the fuzzy 2-sphere
and gives a discrete non-commutative algebra of functions
on ${\bf CP}^{N-1}_F$, represented by matrix multiplication. The matrices are restricted
to ones whose dimension is that of the totally symmetric representations
of $SU(N)$. In the limit of infinite
dimensional matrices
we recover the commutative algebra of functions on ${\bf CP}^{N-1}$.
Derivatives on ${\bf CP}^{N-1}_F$ are also expressed as matrix commutators.
\vfill\eject
\centerline{\bf\S \autosec Introduction}
\bigskip

The concept of non-commutative geometry, [\autorefb
\newcount\Connes\Connes=\refno,\autorefb],
\newcount\Madore\Madore=\refno 
is emerging as one of the most promising 
and interesting new tools in quantum field theory.
It is also providing novel insights into the possible 
space-time structure at the level of quantum gravity.
In quantum field theory it can provide a regularisation
technique which is completely compatible with the space-time
symmetries of the theory,
\hbox{[ \autorefb --\global\newcount\European\global\European=\refno\autorefs
\global\newcount\Presnajder\global\Presnajder=\refno\autorefs
\global\newcount\gropre\global\gropre=\refno\autorefs
\global\newcount\grklpra\global\grklpra=\refno\autorefs
\global\newcount\grklprb\global\grklprb=\refno\autorefs
\global\newcount\grklprc\global\grklprc=\refno\autorefs
\global\newcount\pr\global\pr=\refno\autorefs
\global\newcount\watamuraa\global\watamuraa=\refno\autorefs
\global\newcount\watamurab\global\watamurab=\refno\autorefs
\global\newcount\bbiv\global\bbiv=\refno\autorefs
\global\newcount\Bala\global\Bala=\refno\autorefs
\global\newcount\grost\global\grost=\refno\autorefs
\global\newcount\Balb\global\Balb=\refno\autorefs
\global\newcount\bdmo\global\bdmo=\refno
\kern -3pt \autorefb]},
\newcount\UG\UG=\refno
while in quantum gravity it points the way to radical approaches.
It has also found several applications in string theory\autoref.
\newcount\strings\strings=\refno
In its matrix model or `fuzzy' 
form\footnote{$^\dagger$}{Fuzzy spaces are discrete matrix approximations to continuous
manifolds.}
it promises a radical
alternative to lattice field theory, where problems such 
as chiral fermion doubling are readily avoided [\the\Bala].
A major obstacle to the development of this fuzzy alternative to 
lattice theories is the paucity of fuzzy spaces with explicit descriptions.

An important ingredient in understanding the continuum 
limit of these fuzzy models is the $*$-product. This is 
a non-commutative product for functions that, in the case of fuzzy 
spaces, represents the matrix product.
An explicit example of a $*$-product is known for the fuzzy 2-sphere [\the\Presnajder]. 
It is known that a $*$-product can be defined as a formal power 
series on any manifold that admits a symplectic or 
Poisson structure \hbox{[ \autorefb
\global\newcount\Kontsevich\global\Kontsevich=\refno, \autorefb]},
\global\newcount\rieffel\global\rieffel=\refno 
but few explicit examples are known.

In this paper we present an explicit construction of a $*$-product on
the fuzzy complex projective space $\CPNF$.
While a non-commutative $*$-product on the continuum $\CPN$ is known,
in an integral representation
(see e.g Perelomov\autoref\newcount\ksp\ksp=\refno), to our knowledge this is the first time
an expression for a $*$-product on the fuzzy $\CPNF$ has been given.  
The construction presented here is a generalisation of the construction
of the $*$-product on the 2-sphere given in [\the\Presnajder].

The layout of the paper is as follows: in the next section we give
a brief discussion of harmonic expansions of functions on fuzzy spaces,
by way of motivation for $*$-products and their use in quantum field
theory; in section 3 we give a general discussion of $*$-products 
analysing when they can be expected to exist and, in particular, when
the given construction, based on equivariant products, should exist;
sections 4 and 5 give a description of $\CPN$ in terms of global
co-ordinates; in section 6 the
$*$-product on fuzzy $\CPNF$ is constructed in terms
of projectors and section 7 describes the relation between derivatives
in the continuum and their discrete representation on $\CPNF$;
finally section 8 summarises the conclusions.  Some technical results
required for the main text are reserved for the appendices.
\bigskip
\equno=0
\centerline{\bf \S \autosec Fuzzy Functions}
\bigskip
If one attempts to discretise field theory on a continuous manifold
there are immediate
problems that have to be overcome. Not least is the fact that
continuum symmetries are lost and great care must be
exercised in ensuring that they are recovered again when the continuum
limit is taken. Another problem, which occurs in Fourier space and is
not often remarked upon because the resolution appears to be so
simple, is that the algebra of functions in truncated Fourier space
does not close in general.  For example if one Fourier analyses
functions on a circle,
$$f(\theta)=\sum_{n=-\infty}^{\infty}f_n e^{in\theta},\autoeq$$
and approximates them by cutting off
the Fourier series at some maximum frequency, $L$,
$$f_{L}(\theta)=\sum_{n=-L}^{L}f_n e^{in\theta}\autoeq$$
then the product of two such functions will in general extend
to frequencies up to $2L$ and so the algebra of truncated
functions does not close under multiplication.  
The same problem manifests itself when functions are expanded on
a sphere in terms of spherical harmonics and then approximated
simply by cutting off the expansion at some maximum angular
momentum.  An obvious na{\rm\"\i}ve remedy
is to project after multiplying and just throw
away all the frequencies higher than $L$.
While this brute force method may work, it is not without its
problems --- for example such a process is non-associative in general.
There are sometimes situations
where a more elegant method presents itself which at the same
time does less violence to the group representation theory and 
allows certain spaces to be discretised while preserving their continuum
symmetries.  One approach is to identify the coefficients in an
harmonic expansion with elements of a matrix. If the multiplication
of two functions can be implemented by matrix multiplication then
the matrix algebra will close and no projection is necessary.

Consider for example a 2-dimensional sphere which can be written
as the coset space $S^2\cong SU(2)/U(1)$.
A general function on $SU(2)$
can be expanded in terms of $D$-matrices,
$$f=\sum_{l=0,{1\over 2},1,\ldots}^{\infty}\sum_{m,m^\prime =-l}^{l}
f^l_{m,m^\prime}D^l_{m,m^\prime}.\autoeq$$
To restrict this to a function on $S^2$ the expansion must be restricted
to entries of the $D$-matrices (or linear combinations of them)
which are invariant under the right action
of $U(1)$. The only such entries have $m^\prime=0$,
and hence have integral $l$, since  $m^\prime$ is the $U(1)$ quantum number. 
The $D$-matrices can be constructed so that $D^l_{m,0}$ are independent of the
third Euler angle on $SU(2)$, then they depend only on the polar and azimuthal
angles on $S^2$ and
they are essentially the spherical harmonics --- in standard notation 
$D^l_{m,0}=\sqrt{4\pi\over 2l+1}(-1)^mY^l_{-m}$.
Now the representation theory of $SU(2)$ allows a re-arrangement of
the coefficients in  a truncated expansion
$$f_{L}(\theta,\phi)=\sum_{l=0}^{L}\sum_{m=-l}^{l}
f^l_{m}D^l_{m,0}(\theta,\phi)\autoeq$$
\newcount\SphericalHarmonic\SphericalHarmonic=\equno
into a square matrix. For any given value of $l$,
$\sum_{m=-l}^{m=l}f^l_{m}D^l_{m,0}$ is 
just one component of the row vector obtained from the right
action of an element of $SU(2)$ on the row vector with components $f^l_m$, $m=-l,\ldots,l$.  
For a fixed $l$ 
the row vectors with components $f^l_{m}$ carry an irreducible representation of $SU(2)$.
The set of all coefficients in the expansion (\the\secno.\the\SphericalHarmonic) 
therefore constitute a reducible
representation. 
For example if $L$=1 the number of coefficients $f^l_m$ is 
$1+3=2 \times 2$,
if $L=2$ the number is $1+3+5=3\times 3$ and so on.
For general $L$ the number of terms in this expansion at each value of $l$ is
$2l+1$ giving a total of
$$(L+1)^2=1+3+5+\cdots +(2L+1)\autoeq$$
coefficients, which are in the reducible $(L+1)\times (L+1)$
representation of $SU(2)$. Multiplication of two functions truncated
at the same value of $L$ can now be defined as multiplication
of their associated $(L+1)\times (L+1)$ matrices and 
group representation theory ensures that the resulting product, being
itself a $(L+1)\times (L+1)$ matrix, 
only entails angular momentum up to $L$. 
These matrices define the fuzzy sphere and this matrix 
multiplication induces the $*$-product on the fuzzy sphere.  
It is a non-commutative associative product
and it will be shown later that it reduces to the familiar commutative
product of functions in the limit $L\rightarrow\infty$.
\smallskip
The 2-sphere is rather special in that $SU(2)$ has irreducible representations
for every integer and so matrices of any size can be used to approximate
functions, but more general coset spaces are more restrictive.
Consider, for example, ${\bf CP}^2\cong SU(3)/U(2)$.
Again a function on $SU(3)$ can be expanded in terms of the
representation matrices of $SU(3)$,
$$f=\sum_{l_1,l_2}\sum_{I,I_3,Y;I^\prime,I^\prime_3,Y^\prime}
f^{(l_1,l_2)}_{I,I_3,Y;I^\prime,I^\prime_3,Y^\prime}
D^{(l_1,l_2)}_{I,I_3,Y;I^\prime,I^\prime_3,Y^\prime},\autoeq$$
where the integers $l_1$ and $l_2$ label the irreducible representations
of $SU(3)$ 
($l_1$ and $l_2$ are respectively
the number of symmetric ${\bf 3}$'s and the number of symmetric 
${\bf \bar 3}$'s in the Young diagram
of the representation) and
$I$, $I_3$ and $Y$ are the isospin, third component of isospin and
hypercharge respectively of the little group $U(2)$ (these can be
used to label the weights of any irreducible representation of $SU(3)$ 
unambiguously). 
To describe a scalar function on ${\bf CP}^2$ we must pick out the
parts of the $SU(3)$ representation matrices that are $U(2)$ singlets
under right multiplication. This immediately eliminates all the complex
 representations of $SU(3)$: the ${\bf 3}$, ${\bf \bar 3}$, ${\bf 6}$,
${\bf \bar 6}$ etc.  The remaining real representations require $l_1=l_2=l$
and have dimension $(l+1)^3$.
Again of these  only one column of each representation matrix
survives --- the one given by $I^\prime=I^\prime_3=Y^\prime=0$.
The column vectors ${\cal Y}^{(l,l)}_{I,I_3,Y}:=D^{(l,l)}_{I,I_3,Y;0,0,0}$
thus constitute generalised spherical harmonics on ${\bf CP}^2$
and functions can be expanded as
$$f=\sum_{l}\sum_{I,I_3,Y}f^{(l,l)}_{I,I_3,Y}
{\cal Y}^{(l,l)}_{I,I_3,Y}.\autoeq$$
Again the coefficients fall into  representations of $SU(3)$:
$${\bf 1}+{\bf 8}+{\bf 27}+{\bf 64}+ \cdots.\autoeq$$
Truncating at some maximum value, $L$, of $l$ always allows the
number of coefficients to be arranged in a square matrix: thus
$L=1$ gives ${\bf\bar 3}\times{\bf 3}={\bf 1}+{\bf 8}$;
$L=2$ gives ${\bf\bar 6}\times{\bf 6}={\bf 1}+{\bf 8}+{\bf 27}$;
$L=3$ gives ${\bf\overline {10}}\times{\bf 10}={\bf 1}+{\bf 8}+{\bf 27}+{\bf 64}$;
and so on. Truncating at $L$ results in square matrices of size
${(L+2)(L+1)/2}$, which is the dimension of the symmetric
tensor product of $L$ ${\bf 3}$'s (or $L$ ${\bf\bar 3}$'s),
and 
$$\sum_{l=0}^{L}(l+1)^3 = {(L+2)^2(L+1)}^2/4.\autoeq$$
Again the group representation theory ensures that matrix multiplication
keeps within the same representations and never goes above $L$.

This construction generalises to the higher 
dimensional complex projective spaces ${\bf CP}^{N-1}$ 
where the matrices at 
the smallest non-trivial approximation begin with 
${\bf\overline{N}}\times {\bf N}={\bf 1}+{\bf (N^2-1)}$, the next
being
${{\overline{\bf {N(N+1)}\over 2}}}\times {\bf {N(N+1)\over 2}}={\bf 1}+
\bf{(N^2-1)}+{\bf {N^2(N^2+2N-3)\over4} }$ 
etc. Truncating at $L$ gives a $\left[{(N-1+L)!\over (N-1)!L!}\right]\times\left[{(N-1+L)!\over (N-1)!L!}\right]$ 
matrix representation approximation of ${\bf CP}^{N-1}$.
A similar truncation works for unitary 
Grassmannian manifolds, [\the\UG]. However 
it is not always the case that the representation theory allows
the expansion of a function on a coset space to be described in
terms of square matrices like this. When it can be done we can define
a $*$-product on a fuzzy version of the space.
\bigskip
\equno=0
\centerline{\bf\S \autosec On $*$-products}
\bigskip

In this section we present a general discussion of $*$-products
with emphasis on ``equivariant'' $*$-products.

Suppose we have an algebra $\hat{\cal A}$ of linear operators on a
finite dimensional vector space.  We assume that, if $\hat F\in\A$
then its Hermitian conjugate $\hat F^\dagger\in\A$, so that $\A$ is a
*-algebra.  Let a connected compact Lie group ${\cal G}=\{ g\} $ act
on $\hat{\cal A}$ by adjoint action of unitary matrices:
$$
\hat{F} \mapsto \hat{D}(g) \ \hat{F} \ \hat{D}^{-1}(g), 
\qquad \hat D^\dagger(g)D(g)=\bf 1 .\autoeq $$
\newcount\adjaction\adjaction=\equno
\newcount\adjactionS\adjactionS=\secno
We can assume, by Wedderburn's theorem,\autoref\newcount\Wedderburn\Wedderburn=\refno, 
that $\A$ is the direct sum of full matrix algebras, $Mat_d$, of
$d\times d$ matrices: $\A=\bigoplus_{d} Mat_d$. As the $\hat{D}(g)$
action preserves $\A$, it also decomposes as $\hat{D}(g)=\bigoplus_{d}
\hat{D}^{(d)}(g)$.
Since  $Mat_d$ is simple, the two-sided ideals of $\A$ are all
direct sums of some of the $Mat_d$ or just $\{ 0 \} $.

To get a $*$-product we need, in addition, a function $\hat{\rho}^*$ on a
manifold ${\M}$ with values in $\A^*$, the dual of $\A$. Then,
$<\hat\rho^*,\hat{F}>:=F$ is a function on $\M$: 
$$
<\hat\rho^*,\hat{F}>(\xi)\equiv <\hat\rho^*(\xi),\hat{F}>=F(\xi),\autoeq
$$
where $\xi\in \M$.
This map $\A \rightarrow {\cal C}_F^\infty ({\cal M})\subset{\cal C}^\infty ({\cal M})$ (assuming
appropriate continuity requirements) induces an algebra structure on ${\cal
C}^\infty _F ({\cal M})$ if its kernel, $Ker$, is a two-sided ideal in
$\A$, that 
is if $Ker$ is a direct sum of some of the $Mat_d$ or $\{ 0 \}$. If
that is the case, ${\cal C}_F^\infty ({\cal M})\cong \A/Ker$, and its
algebra product is defined by 
$$(F*G)(\xi)=<\hat\rho^*(\xi),\hat{F}\hat{G}>,\autoeq$$
\newcount\rhostarprod\rhostarprod=\equno
\newcount\rhostarprodS\rhostarprodS=\secno
where $\hat F, \hat G \in \A$.

The action (\the\secno.\the\adjaction) on $\A$ induces an action on
its dual $\A^*$ which we denote by $\hat{F}^* \rightarrow \hat{D}^*
(g)^{-1} \hat{F}^* \hat{D}^* (g)$:
$$
<\hat{F}^*, \hat{D} (g) \hat{F} \hat{D} (g)^{-1}>=<\hat{D} (g)^{*-1}
\hat{F}^* \hat{D} (g)^{*},\hat{F}>.\autoeq$$

Until now there is no requirement that $\hat{\rho}^*(\xi)$ is a state or
has equivariance. The setting is very general. Suppose we now ask that 
$\hat{\rho}(\xi)$ is a state:
$$
<\hat\rho^*(\xi),\hat{F}^\dagger>=\overline{<\hat\rho^*(\xi),\hat{F}>},\autoeq$$
\newcount\hermit\hermit=\equno
$$<\hat\rho^*(\xi),\hat{F}^\dagger \hat{F}>\  \geq \ 0, \autoeq$$
\newcount\posit\posit=\equno
$$<\hat\rho^*(\xi),\hat{\bf 1}>=1, \autoeq$$
\newcount\norm\norm=\equno
where bar denotes complex conjugation.
Then $\hat{\rho}^*(\xi)$ can be identified with a density matrix
$\hat\rho(\xi)$ by setting 
$$
<\hat\rho^*(\xi),\hat{F}>=\Tr (\hat\rho(\xi)\hat{F}).\autoeq$$
\newcount\densitymatrix\densitymatrix=\equno
\newcount\densitymatrixS\densitymatrixS=\secno
For equivariance we assume that $g$ acts transitively on $\M$,
$\xi\rightarrow g\xi$, such that
$$
\hat{\rho}^* (g\xi)=\hat{D}^*(g)\ \hat{\rho}^* (\xi)\
\hat{D}^*(g^{-1}).\autoeq $$
\newcount\correctionbal\correctionbal=\equno
\newcount\correctionbalS\correctionbalS=\secno

Now each $Mat_d$ and $\A$ can be decomposed into irreducible
tensor operators: 
$$
\eqalignno{Mat_d &=\hbox{Span} \{ \hat{T}^{(l)}_M (d)\} , \cr
\hat{D}^{(d)}(g) \ \hat{T}^{(l)}_M (d) \ \hat{D}^{(d)}(g)^{-1} & =
\sum_{M'}\hat{T}^{(l)}_{M'}(d) D^{(l)}_{M'M} (g) ,\autoeql}$$
where $g\mapsto D^{(l)}(g)$ is a unitary 
irreducible representation. Let $\{ \hat{T} ^{* (l)}_M (d)\} $ 
be the dual basis: 
$$
<\hat{T} ^{*(l')}_{M'} (d'),\hat{T}^{(l)}_M (d)>=\delta_{ll'}\delta
_{dd'} \delta_{MM'}.\autoeq$$
It transforms as 
$$\hat{D}^{*(d)}(g^{-1}) \ \hat{T}^{*(l)}_M (d) \ \hat{D}^{*(d)}(g) =
\sum_{M'}\hat{T}^{*(l)}_{M'}(d) \ D^{*(l)}_{M'M} (g^{-1}) ,\autoeq$$
where
$$\hat{D}^{*(l)}_{M'M}(g)\ \hat{D}^{(l)}_{M'N}(g) =
\delta_{MN}.\autoeq 
$$

We can expand 
$$\hat{\rho}^* =\sum_{d,l,M} \rho_M^{(l,d)} \hat{T}^{*(l)}_M (d):=
\sum_{l,d} \hat{\rho}^{*(l,d)},$$ 
where
$$\quad\rho_M^{(l,d)}:\M\rightarrow {\bf C}\quad
\hbox{and}\quad\hat{\rho}^{*(l,d)}=\sum_M \rho_M^{(l,d)}
\hat{T}^{*(l)}_M (d) .\autoeq$$ 

Now Wedderburn's theorem implies that
for a $*$-product to exist, either {\it all} functions $\rho_M^{(l,d)}$
for a fixed $d$, or {\it none}, must be zero, because if $\hat\rho^*$
has a kernel consistency requires that it be a full matrix algebra.
In fact, because of equivariance, we shall now show that it is
sufficient to check if $\hat\rho^{*(l,d)}$ is zero or not at one
point, which we shall call the origin and denote by $\xi_o$. We have 
$$
\hat{\rho}^{*(l,d)} (g\xi_o)=\hat{D}^{*(d)}(g) \ \hat{\rho}^{*(l,d)}
(\xi_o) \ \hat{D}^{*(d)}(g^{-1}),\autoeq$$ 
or
$$\eqalignno{\sum_M \rho_M^{(l,d)}(g\xi_o) \ \hat{T}_M^{*(l)} (d) & = 
\sum_{M,M'} \rho_M^{(l,d)}(\xi_o) \ \hat{T}_{M'}^{*(l)}(d) \ D^{(l)}_{M'M}(g) \cr
\Rightarrow \rho_{M'}^{(l,d)}(g\xi_o) & = D^{(l)}_{M'M}(g) \rho_M^{(l,d)}
(\xi_o) .\autoeql}$$
So, from equivariance, 
$$
\hat{\rho}^{*(l,d)}(\xi_o)=0 \Rightarrow \rho_M^{(l,d)}(\xi_o)=0 \quad\hbox{and}\quad
\hat{\rho}^{*(l,d)}= 0.\autoeq$$
Thus, with equivariance, it is enough to check that
$\hat{\rho}^{*(l,d)}(\xi_o)=0$, either for all $l$ or no $l$, for each fixed
$d$, to verify if $*$ exists.

\bigskip
We remark that it is not necessary to assume (\the\secno.\the\norm) separately,
as we can arrange to have it with (\the\secno.\the\hermit) and
(\the\secno.\the\posit):
$$
<\hat\rho^*(g\xi_o),\hat{\bf 1}>=<\hat{\rho}^* (\xi_o),\hat{D}(g) \ \hat{\bf 1} \ 
\hat{D} (g^{-1})>=<\hat\rho^*(\xi_o),\hat{\bf 1}>=<\hat\rho^*(\xi_o),
\hat{\bf 1}^{\dag} \hat{\bf 1}> \autoeq$$
which, by (\the\secno.\the\posit), is a constant non-negative number, $c$. As the ideal containing
$\hat{\bf 1}$ is $\A$, $c$ cannot be zero if there is a
non-trivial $*$-product. So we can work with $\hat\rho^*/c$
instead so that (\the\secno.\the\norm) is enforced. 
As for (\the\secno.\the\hermit) and (\the\secno.\the\posit),
they are natural. Eq. (\the\secno.\the\hermit) gives real functions for
hermitian operators, and (\the\secno.\the\posit) gives $\overline{F}*
F(\xi)\geq 0.$\hfill\break

Note that if functions on $\M$ do not carry an IRR $l$ with the
correct multiplicity, it can happen that $\A$ admits no $*$-product.
This problem occurs, for example, if $\A$ is the $8\times 8$ matrix
algebra and $\hat\rho(\xi_o)$ is $a{\bf 1}+b[ad(Y)]$ (where $ad(Y)$ is
the adjoint generator of hypercharge) with $a$ and $b$ chosen so that
(\the\secno.\the\hermit) and (\the\secno.\the\posit) are
satisfied. Then, $\hat\rho$ gives a map to functions on ${\bf
CP}^2$. The latter has $8$ only once, but $\A$ has two $8$'s, so there
is no $*$-product (for a general discussion
see\autoref).\newcount\Mukunda\Mukunda=\refno\ A $*$-product does
however exist on ${\bf CP}^2$, for suitable $\hat\rho$, which we
construct later. 

It is useful to note the following. Quite generally, in the
equivariant case, with $t_A$ an orthonormal basis (in the trace norm)
for the Lie algebra of $G$, 
$$\Tr (\hat\rho (g\xi_o) t_A)=\xi_A (g)=(\hbox{Ad}g)_{AB} \xi_B (1).\autoeq$$
\newcount\adjxi\adjxi=\equno
\newcount\adjxiS\adjxiS=\secno
Writing $\hat\rho(g)=(\sum \eta_B (g)t_B+$ terms orthogonal to $t_B$),
only the first term survives the tracing, so that $\eta_A=\xi_A$, with
$t_A$ normalised appropriately. $\xi_A$ maps $G/H$ to an adjoint orbit
and provides coordinate functions on $G/H$.

To escape the limitation of only getting $*$-star products on adjoint
orbits, we may have to modify the requirement of equivariance.

In the subsequent construction of a $*$-product on $\CPN$ we shall
restrict our considerations to the case where $\hat\rho$ is a rank $1$
projector and we shall use the notation $\P$ for $\hat\rho$ (or
$\P_L=\hat\rho$ for its $L$-fold symmetric product, as explained
later).

\bigskip
\equno=0
\centerline{\bf\S \autosec Global Co-ordinates on $\CPN$}
\bigskip
We now turn to an explicit construction of the complex
projective space ${\bf CP}^{N-1}$, which can be defined as the space of 
vectors of unit norm in ${\bf C}^N$ modulo the phase. Since a unit vector 
$|\psi>$ up to a phase defines a projection operator $\P:=|\psi><\psi|$, 
an equivalent definition for ${\bf CP}^{N-1}$ is as the space of all 
projection operators of rank one on ${\bf C}^N$, i.e.,
$${\bf CP}^{N-1}:= \{\P\in Mat_N;\,\,\P^\dagger=\P, \P^2=\P,
\Tr\P=1\}.\autoeq$$ 
\newcount\definitioncpn\definitioncpn=\equno
\newcount\definitioncpnS\definitioncpnS=\secno
To construct a set of global coordinates for 
$\CPN$, we choose a basis for $N\times N$ 
hermitian matrices $\{ {t_\hA} \}$, $\hA=0,\cdots,N^2-1$, consisting
of $t_0={\bf 1}/\sqrt{N}$ and $\{t_A\}$, $A=1,\cdots, N^2-1$, forming
an orthogonal basis for the Lie algebra of $SU(N)$. We will normalize
them by requiring  
$$\Tr ({t_\hA} {t_\hB})= \delta_{\hA\hB}.\autoeq$$
\newcount\normalizeS\normalizeS=\secno
\newcount\normalize\normalize=\equno
This requirement implies that $t_0={1\over\sqrt{N}}{\bf 1}$ and 
$t_A$'s are related to the Gell-Mann matrices $\lambda_A$ by 
$t_A=\lambda_A/\sqrt{2}$. Thus, 
$$t_At_B={1\over N}\delta_{AB}{\bf 1} + 
{1\over\sqrt{2}}\bigl({d_{AB}}^C + i{f_{AB}}^C\bigr)t_C,\autoeq$$
\newcount\GMalgebraS\GMalgebraS=\secno
\newcount\GMalgebra\GMalgebra=\equno
where ${f_{AB}}^C$ and ${d_{AB}}^C$ are, respectively, the structure constants 
and the components of the symmetric invariant tensor of $SU(N)$ in the 
Gell-Mann basis. The $d$-tensor is traceless on each pair of
indices. For raising and lowering indices we will use the Kronecker $\delta$.

Expanding $\P$ in terms of the basis,
$$\P=\xi^\hA t_\hA= \xi^0t_0+ \xi^At_A.\autoeq$$
The condition that $\P$ is a rank one projection operator leads to the 
following conditions on the coordinates $\xi^\hA$, 
$$\xi^0={1\over\sqrt{N}},\qquad\xi^A\xi_A={N-1\over N},
\qquad {d_{AB}}^C\xi^A\xi^B={\sqrt{2}(N-2)\over N}\xi^C.\autoeq$$
\newcount\tl\tl=\equno
\newcount\tlS\tlS=\secno
These form a set of quadratic constraints which describe $\CPN$ embedded
in the $N^2$-dimensional Euclidean space ${\bf R}^{N^2}$, or in 
${\bf R}^{N^2-1}$ since $\xi^0$ is a fixed constant.
For example, for $N=2$ we have $A,B=1,2,3$ and the above equations reduce to 
that of a sphere, or ${\bf CP}^1$, of radius 
$1/\sqrt{2}$ embedded in ${\bf R}^3$ because the 
$d$-tensor vanishes for $SU(2)$.

The coordinates for $\CPN$ can be constructed easily by noting that any 
$\P\in\CPN$ can be obtained from an arbitrarily chosen origin 
$\P_o$ by rotating it with $g\in U(N)$, $\P=g\P_og^\dagger$. Of course there 
is no unique element $g$ associated with $\P$. In fact, any two
elements of $U(N)$ that are related by $g^\prime=gh$ where 
$h\in U(1)\times U(N-1)$
give rise to the same point of $\CPN$, as can be seen by going to the basis 
of ${\bf C}^N$ in which $P_o$ is diagonal. (This leads to still another 
characterization of the 
complex projective space, i.e., $\CPN=U(N)/[U(1)\times U(N-1)]$.)
Using this fact one can obtain coordinates, $\xi^\hA$,
corresponding to an arbitrary point of $\CPN$,
$\P=g\P_og^\dagger$, from the coordinates $\xi_o^\hA$ of the origin
$\P_o$ 
as follows: 
$$\xi^\hA=\Tr (\P t^\hA)=\Tr (g\P_og^\dagger t^\hA)=\xi_o^\hB
\Tr (g t_\hB g^\dagger t^\hA)={(Ad(g))^{\hat A}}_{\hat B}\xi^{\hat B}_o,\autoeq$$
\newcount\fiducialprojector\fiducialprojector=\equno
so that $\xi^{\hat A}$ map $\CPN$ to an adjoint orbit of $U(N)$
fulfilling (\the\adjxiS.\the\adjxi). 

It is important for what follows that $\P$ fulfills the property
(\the\correctionbalS.\the\correctionbal):
$$
g^{-1}\xi^At_A g=(Ad(g))_{AB} \xi_B t_A. \autoeq $$
Here $g\in U(N)$ and $g\rightarrow Ad(g)$ defines its adjoint
representation.

\bigskip
\equno=0
\centerline{\bf\S \autosec The Geometry of $\CPN$}
\bigskip

The coordinates $\xi^\hA$, $\hA=0,\ldots,N^2-1$ constitute an over-complete,
but globally well-defined, coordinate system for $\CPN$. It is 
therefore useful to use them to describe geometrical structures on $\CPN$ 
such as Fubini-Study metric and K\"ahler structure. To this end let us regard
$\CPN$ as a manifold embedded in the space ${\bf R}^{N^2}$ of all
hermitian $N\times N$ matrices. At a given point $\P\in\CPN$ we
can decompose ${\bf R}^{N^2}$ into the subspace   
${\rm T}_\P\CPN$ consisting of vectors tangential to $\CPN$ and its 
orthogonal complement. Since the action of $U(N)$ spans all directions
tangential to $\CPN$ at $\P$, and $\P$ is rotated by the adjoint
action of $U(N)$, any vector in ${\rm T}_\P\CPN$ 
must be of the form,
$$\T=i\,Ad(\Lambda)\P=i[\Lambda,{\cal P}] \autoeq$$
\newcount\tangentvectortwo\tangentvectortwo=\equno
for some hermitian matrix $\Lambda$. 
This immediately implies that $\T$ must satisfy
$${\T}^\dagger=\T, \quad \{\P,\T\}=\T, \quad \Tr\T=0.\autoeq$$
\newcount\tangentvector\tangentvector=\equno

Note that if $\Lambda$ is a generator of the stability 
subgroup $U(1)\times U(N-1)$, the RHS of
(\the\secno.\the\tangentvectortwo) vanishes so that vectors $\T$ span
a vector space of dimension of $N^2-(N-1)^2-1=2N-2$, which agrees
with the dimension of $\CPN$. 

The vectors in the orthogonal complement of ${\rm T}_\P\CPN$, on the other 
hand, can be represented by the generators $\N$ of the stability subgroup 
of $U(N)$. They satisfy
$$[\P,\N]=0.\autoeq$$\newcount\commN\commN=\equno
One can see this by noting that all such vectors are orthogonal to 
$\T=i[\Lambda, \P]$,
$$\Tr(\N\T)= i\Tr(\N[\Lambda, \P])=i\Tr([\P,\N]\Lambda)=0.\autoeq$$ 

These facts are now used to describe the K\"ahler structure on $\CPN$. The 
 K\"ahler structure consists of the following three mutually compatible 
structures:

1) {\it Complex structure}: for any hermitian matrix $\M$, regarded as a vector 
at $\P$, define 
$$J(\M):=-i[\P,\M].\autoeq$$
\newcount\complex\complex=\equno
\newcount\complexS\complexS=\secno
If $\M$ is normal to $\CPN$, i.e., if $\M=\N$, then $J(\N)=0$ trivially. 
If $\M$ is tangential, i.e., $\M=\T$, then
$$J^2(\T)=-[\P,[\P,\T]]=-\P(\P\T -\T\P)+ (\P\T -\T\P)\P
=-\P\T-\T\P+2\P\T\P=-\T,\autoeq$$\newcount\Jsquared\Jsquared=\equno
where in the last step we have used (\the\secno.\the\tangentvector) and 
$\P\T\P=0$ 
which follows immediately from that equation. Therefore, $J$ is a
complex structure on $\CPN$. In view of with (\the\secno.\the\commN)
and (\the\secno.\the\complex), equation (\the\secno.\the\Jsquared)
shows that $-J^2$ is a projector to the tangent space of $\CPN$. 

2) {\it Metric}: for two hermitian matrices $\M_1$ and $\M_2$ define  
$${\cal G}(\M_1,\M_2):= \Tr\big(-J^2(\M_1)\, \M_2\big)=-\Tr
([P,\M_1][P,\M_2]).\autoeq$$ 
\newcount\metric\metric=\equno
\newcount\metricS\metricS=\secno
This vanishes if any one of the arguments is a normal vector and on
tangent vectors it agrees with the trace metric.  It is the metric on
$\CPN$ induced from the trace metric for hermitian matrices.  One can
show that ${\cal G}(J(\M_1),J(\M_2))={\cal G}(\M_1,\M_2)$.

3) {\it Symplectic structure}: for two matrices $\M_1$ and $\M_2$, 
define an antisymmetric two-form $\Omega$ by 
$$\Omega(\M_1,\M_2):=\G(J(\M_1),\M_2)=-i\Tr(\P[\M_1,\M_2]).\autoeq$$
\newcount\symplectic\symplectic=\equno   
\newcount\symplecticS\symplecticS=\secno   
It vanishes if any of the arguments is normal to $\CPN$. Thus, it is a 
two-form on $\CPN$. It is in fact closed, as we shall show  
in section 7.

One can  
combine ${\cal G}$ and $\Omega$ to obtain a tensor $K$ on $\CPN$,
$$K:={1\over2}({\cal G}+i\Omega),\autoeq$$
\newcount\KGOmega\KGOmega=\equno
\newcount\KGOmegaS\KGOmegaS=\secno
and it is a straightforward exercise to show that 
$$K(\M_1,\M_2)=\Tr(\P\M_1\M_2)-\Tr(\P\M_1 \P\M_2)=\Tr[\P\M_1({\bf
1}-\P)\M_2].\autoeq$$ 
\newcount\genKdef\genKdef=\equno
\newcount\genKdefS\genKdefS=\secno

The construction of the K\"ahler structure described here also holds
for other spaces of projection operators of a fixed rank, i.e. unitary
Grassmannian manifolds. However, the fact that $\CPN$ consists of rank
one projection operators further simplifies the above equation to
$$K(\M_1,\M_2):=\Tr(\P\M_1\M_2)-\Tr(\P\M_1)\Tr(\P\M_2).\autoeq$$
\newcount\cpnKdef\cpnKdef=\equno
\newcount\cpnKdefS\cpnKdefS=\secno
This form of $K$ will be used crucially in the construction 
of fuzzy $\CPN$ in the following section. In terms of the components 
with respect to the basis $t_A$ ($t_0$ components all vanish) one finds
$$K_{AB}:= K(t_A,t_B)={1\over N}\delta_{AB}+{1\over\sqrt{2}}({d_{AB}}
^C+if_{AB}{}^C)\xi_C-\xi_A\xi_B,
\autoeq$$
$${\cal G}_{AB}=2\hbox{Re}K_{AB},\quad\quad
\Omega_{AB}=2\hbox{Im}K_{AB},\quad\quad
J^A{}_B=\delta^{AC}\Omega_{CB}.
\autoeq$$ 
Because of our normalisation of the matrices $t_A$,
(\the\normalizeS.\the\normalize), the indices $A,B,\ldots$ are raised
and lowered with $\delta^{AB}$ and $\delta_{AB}$ respectively.  It is
shown in the appendix that ${P^A}_B:=\delta^{AC}{\cal G}_{CB}$ is a
rank $2(N-1)$ projector and in fact $P=-J^2$. Alternatively,
Eq. (\the\metricS.\the\metric) will yield that result directly by
splitting and combining traces containing the one-dimensional
projector $P$ as in (\the\cpnKdefS.\the\cpnKdef).  In future we shall
not distinguish between ${\cal G}$ and $P$, nor between $\Omega$ and
$J$, and shall write
$$K={1\over 2}(P+iJ),\autoeq$$
\newcount\Kdef\Kdef=\equno
\newcount\KdefS\KdefS=\secno
with
$$P_{AB}={2\over N}\delta_{AB}+{\sqrt{2}}({d_{AB}}^C\xi_C)-2\xi_A\xi_B\autoeq$$
and 
$$J_{AB}=\sqrt{2}{f_{AB}}^C\xi_C.\autoeq$$
In fact, as shown in the appendix, $K$ itself is a rank $N-1$
projector --- it can be interpreted as a projector from the redundant,
global coordinates $\xi_A$ to local (anti-)holomorphic coordinates on
$\CPN$. That $K$ is a projector is obvious from
(\the\KdefS.\the\Kdef), $J^2=-P$ and $PJ=JP=J$.

\bigskip
\equno=0
\centerline{\bf\S \autosec Fuzzy Complex Projective Spaces}
\bigskip

We now turn to the construction of functions on $\CPN$ and their $*$-product, generalizing the construction given for $S^2_F\cong{\bf CP}^1_F$ 
in [\the\Presnajder]. 
While a non-commutative $*$-product on the continuum ${\bf CP}^{N-1}$ has
been known for some time [\the\ksp],
we construct here a $*$-product on the fuzzy $\CPNF$, with a finite number of
degrees of freedom.

In order to describe the harmonic expansion of functions on $\CPN$ one
only requires representations which are symmetric products of the
fundamental representation of $SU(N)$, i.e. the ${\bf N}$
representation, (or the complex conjugate ${\bf\bar N}$
representation). So the construction starts with an $N$-dimensional
Hilbert space, $\H_1:={\bf N}={\bf C}^N$.  To represent functions at
the level $L$, we use as our Hilbert space , $\H_L$, which is the
$d_L={(N-1+L)!\over(N-1)!L!}$-dimensional irreducible representation
of $SU(N)$ obtained from the $L$-fold symmetric tensor product of
$\H_1$.  Associated with a point $\P$ in $\CPN$ let us consider the
$L$-fold tensor product of $\P$,
$$\P_L:=\P\otimes\cdots\otimes\P.\autoeq$$ 
Being an $L$-fold tensor product of the same operator, $\P_L$ is 
a well-defined 
operator on $\H_L$. Note that $\P_L$ is again a projection operator of 
rank one. We will use this 
property of $\P_L$ later.

With each operator $\hat F$ on ${\cal H}_L$, we construct the corresponding 
function $F_L(\xi)$ on $\CPN$ using the equivariant mapping prescription,
$$F_L(\xi):=\Tr(\P_L(\xi) \hat F).\autoeq$$
\newcount\tracefun\tracefun=\equno
\newcount\tracefunS\tracefunS=\secno 
In this way we define an injective mapping from operators $\hat F$ on
${\cal H}_L$ into functions $F_L$ on $\CPN$ (the injectivity is
actually proved at the end of next section).  The functions $F_L$ are
sufficient to reconstruct the operator $\hat F$.  The target space of
this mapping is derived in section 7, it is what we denote by $\CPNF$
and is isomorphic to the space of $d_L\times d_L$ matrices.

A $*$-product on this space of functions is defined as 
\secno=6
$$(F_L * G_L)(\xi):=\Tr (\P_L\hat F \hat G).\autoeq$$
\newcount\starp\starp=\equno
Associativity of the $*$-product is guaranteed by construction
and derives from the associativity of matrix multiplication. 
Our aim is to derive an explicit, closed expression for the 
$*$-product (\the\secno.\the\starp) (or (\the\rhostarprodS.\the\rhostarprod)), solely in terms of the functions $F_L$ and $G_L$, 
and show that it reduces to the normal
product of two functions in the limit $L\rightarrow\infty$.

At the level $L=1$, the only functions allowed are functions linear in 
$\xi^\hA$. This is because any hermitian operator acting on the fundamental 
representation $\H_1$ of $SU(N)$, can be expanded in terms of $t_\hA$. For 
$\hat F=F^\hA t_\hA$, the corresponding function $F_1(\xi)$ become
\secno=6
$$F_1(\xi)= F^\hA\xi_\hA.\autoeq$$
In particular, $t_\hA$ produces coordinate functions $\xi_\hA$,
$$\xi_\hA=\Tr(\P t_\hA).\autoeq$$
The $*$-product between coordinate functions, $\xi_\hA*\xi_\hB:=\Tr(\P t_\hA t_\hB$)
combined with (\the\cpnKdefS.\the\cpnKdef) yields the following important relation
$$\xi_\hA*\xi_\hB 
=\xi_\hA\xi_\hB + K_{\hA\hB},\autoeq$$
\newcount\xiprod\xiprod=\equno
where $K_{\hA\hB}$ is the hermitian structure. Note that $K_{0\hA}$
vanishes for all $\hA$.

For any finite $L$, functions and their $*$-product are constructed using 
hermitian operators on $\H_L$ according to the prescriptions 
(\the\secno.2), (\the\secno.\the\starp). Given two operators $\hat F$ and $\hat G$
write them in the following form,
$$\hat F= F_{\hA_1\cdots \hA_L}t^{\hA_1}\otimes\cdots\otimes t^{\hA_L},$$
$$\hat G= G_{\hA_1\cdots \hA_L}t^{\hA_1}\otimes\cdots\otimes
t^{\hA_L},\autoeq$$ 
\newcount\correctionbalb\correctionbalb=\equno
where the coefficient tensors are totally symmetric.
Of course, for a given operator on ${\cal H}_L$ there is no unique
expression of the above form. In fact, a choice of symmetric tensor
corresponds to a particular extension of the operator to the whole
tensor product space ${\cal H}_1\otimes {\cal H}_1 \cdots \otimes {\cal
H}_1$. This ambiguity will eventually disappear because the
construction of functions and their $*$-product depend only on
operators acting on $\H_L$. The functions corresponding to
(\the\secno.\the\correctionbalb) are
$$F_L(\xi)= F_{\hA_1\cdots \hA_L}\xi^{\hA_1}\cdots\xi^{\hA_L},$$
$$G_L(\xi)= G_{\hA_1\cdots \hA_L}\xi^{\hA_1}\cdots\xi^{\hA_L},\autoeq$$
and their $*$-product becomes
$$(F_L*G_L)(\xi)=F_{\hA_1\cdots \hA_L}G_{\hB_1\cdots \hB_L}
(\xi^{\hA_1}*\xi^{\hB_1})\cdots(\xi^{\hA_L}*\xi^{\hB_L}).\autoeq$$
\newcount\FGprod\FGprod=\equno
Since $\xi^0={1\over\sqrt{N}}$ is a constant, all functions can be
considered as polynomials in just $\xi^A$ of degree $\le L$. 

Now, in order to express this in the final form, we first substitute the 
relation (\the\secno.\the\xiprod) into the above equation and expand
it in powers of $K^{\hA\hB}$ to get
$$\eqalignno{
(F_L*G_L)(\xi)=& F_L (\xi) G_L (\xi) +\sum_{l=1}^L{L!\over (L-l)!l!}
F_{\hA_1\cdots \hA_l\hA_{l+1}\cdots \hA_L}\xi^{\hA_{l+1}}\cdots\xi^{\hA_L}
G_{\hB_1\cdots \hB_l\hB_{l+1}\cdots \hB_L} \cr
&\qquad\qquad\qquad \xi^{\hB_{l+1}}\cdots\xi^{\hB_L}
K^{\hA_1\hB_1}\cdots K^{\hA_l\hB_l}.\autoeql}$$ 
\newcount\FGKxi\FGKxi=\equno
where the first term is the ordinary commutative product, and will be
integrated into the sum as the $l=0$ term for convenience. Finally,
using the relation 
$$\partial_{\hA_1}\cdots\partial_{\hA_l} F_L(\xi)={L!\over (L-l)!}
F_{\hA_1\cdots \hA_l\hA_{l+1}\cdots \hA_L}\xi^{\hA_{l+1}}\cdots\xi^{\hA_L},
\autoeq $$
and the fact that $K^{0\hA}=0$, we get
$$(F_L*G_L)(\xi)=\sum_{l=0}^L{(L-l)!\over L!l!}
[\partial_{A_1}\cdots\partial_{A_l}F_L(\xi)]K^{A_1B_1}\cdots K^{A_lB_l}
[\partial_{B_1}\cdots\partial_{B_l}G_L(\xi)].\autoeq$$
\newcount\KprodS\KprodS=\secno
\newcount\Kprod\Kprod=\equno
Note again that in arriving at (\the\secno.\the\Kprod) we have
extended functions and derivatives to outside $\CPN$ and finally 
evaluated the result on $\CPN$. However, all this extension should be
regarded as a convenient way of calculation because the final
expression involves functions on $\CPN$ and derivatives along $\CPN$
only, as we will explicitly show below.

Equation (\the\secno.\the\Kprod) is one of the central results of this
paper and generalises the result for $S^2$ derived in
[\the\Presnajder].  Only the $l=0$ term survives in the limit
$L\rightarrow\infty$, which shows that the $*$-product reduces to
ordinary multiplication of functions in the continuum limit, with
corrections being of order $1/L$.  Note that the limit should be taken
with all functions fixed.

As mentioned earlier, and proven in the appendix, the matrix $K_{AB}$
is a projector.  In fact the derivatives in (\the\secno.\the\Kprod),
which are flat in ${\bf R}^{N^2-1}$ are being projected onto the
holomorphic tangent space of $\CPN$ and are actually covariant
derivatives there.  Note that, since $K$ is hermitian, it gives a
holomorphic derivative when acting to the right, as in
$K^{AB}(\partial_BF)$, but an anti-holomorphic derivative when acting
to the left, as in $(\partial_BF)
K^{BA}=\overline{K}^{AB}(\partial_BF)$, where the bar represents
complex conjugation. Thus, if our algebra of functions permitted us to
construct holomorphic or anti-holomorphic functions, the $*$-product
of a (anti-) holomorphic function with another (anti-) holomorphic
function would always reduce to the ordinary product. More generally
the $*$-product, $F_L*G_L$, is an ordinary product if $G$ is
anti-holomorphic regardless of the form of $F$ or, conversely, if $F$
is holomorphic regardless of the form of $G$.  Another point to note
is that the complex structure is reversed, $J\rightarrow -J$, if the
original Hilbert space is identified with the complex conjugate
fundamental representation $\bar {\bf N}$ rather than the ${\bf N}$.

The structure here is perhaps most clearly understood by looking at
the simplest case, $N=2$. Then $P_{AB}=\delta_{AB}-2\xi_A\xi_B$ and
$J_{AB}=\sqrt{2}\epsilon_{ABC}\xi^C$. The constraints imply that
$\xi_A \xi^A=1/2$ and so define a unit vector in ${\bf R}^3$, $n_A
=\sqrt{2}\xi_A$, so that $P_{AB}=\delta_{AB}-n_An_B$ and
$J_{AB}=\epsilon_{ABC}n^C$. Clearly $P=-J^2$ and $P$ is a projector
from ${\bf R}^3$ onto the unit sphere while $J$, when restricted to
${\bf n.n}=1$, represents the complex structure on ${\bf CP}^1$.  In
view of the identity $J^3=-J$, the combination $K=(P+iJ)/2$ is a rank
one projector onto a complex holomorphic coordinate on ${\bf CP}^1$
$(JK=-iK)$. This interpretation survives to higher $N$ also and gives a
geometric interpretation of the $*$-product (\the\secno.\the\Kprod).

In a standard geometrical construction a covariant derivative on a
curved space can be obtained by embedding the space in a flat
Euclidean space of higher dimension and projecting the ordinary
derivative in the Euclidean space onto the tangent space of the curved
manifold. When the Euclidean derivatives are restricted to act on
tensors already projected to the tangent space of the curved manifold,
the projected flat derivative is a covariant derivative.  There is a
simplification in the construction here, because the projector
$K_{AB}$ satisfies [\the\UG]
$$K^{AB}K^{CD}\partial_B K_{DE}=0\autoeq$$\newcount\KKdK\KKdK=\equno
\newcount\oi\oi=\equno
\newcount\oiS\oiS=\secno
which implies that
\secno=6
$$K^{AB}K^{CD}\partial_B\left({K_D}^E\partial_E
F\right)=K^{AB}K^{CD}\partial_B\partial_DF\autoeq$$ 
\newcount\oib\oib=\equno
\newcount\oibS\oibS=\secno
since $K^2=K$, with an obvious generalisation to derivatives acting on
higher rank tensors. 
This identity can be proven
\genKdefS=5
using the last form of $K_{AB}$ in (\the\genKdefS.\the\genKdef),
$K_{AB}=Tr[\P t_A({\bf 1}-\P)t_B]$, and completeness of the matrices
$t_A$.
%
An alternative, more detailed proof, is given in appendix B.

So, defining $\nabla_A:={K_A}^B\partial_B$ and
${\overline\nabla}_A:={{\overline K}_A}^B\partial_B$ and using
(\the\oibS.\the\oib) and its generalization to convert the successive
partial derivative to covariant derivatives in (6.12), the $*$-product
is 
$$(F_L*G_L)(\xi)=\sum_{l=0}^L{(L-l)!\over L!l!}
[\overline\nabla_{A_1}\cdots\overline\nabla_{A_l}F_L(\xi)]K^{A_1B_1}\cdots 
K^{A_lB_l} [\nabla_{B_1}\cdots\nabla_{B_l}G_L(\xi)].\autoeq$$

Converting from global coordinates, $\xi^A$ with $A=1,...,N^2-1$, to
local holomorphic coordinates, $z^i$ with $i=1,...,N-1$ and $z^{\bar
i}:={\bar z}^i$ we have the correspondences
$$K_{AB}\rightarrow {1\over 2}\left(\G_{i\bar j}+i\Omega_{i\bar j}\right)
=i\Omega_{i\bar j},\qquad
K^{AB}\rightarrow {1\over 2}\left(\G^{\bar j i}+i\Omega^{\bar j i}\right)=
i\Omega^{\bar j i},\autoeq $$
where $\G_{i\bar j}$ is the Fubini-Study metric and $\Omega_{i\bar j}$
the K\"ahler 2-form, with $\G_{i\bar j}=\G_{\bar j i}=i\Omega_{i\bar
j}=-i\Omega_{\bar j i}$, and $\Omega^{\bar j i}=\G^{\bar j n}\G^{i\bar
m} \Omega_{n\bar m}$.  Equation (6.15) in local coordinates takes the
from $$
(F_L*G_L)(z,\bar z)=
\sum_{l=0}^L{(L-l)!\over L!l!}
[\nabla_{{\bar j}_1}\cdots\nabla_{{\bar j}_l}F_L(z,\bar z)](i\Omega^{{\bar j}_1 i_1})
\cdots (i\Omega^{{\bar j}_l i_l})
[\nabla_{i_1}\cdots\nabla_{i_l}G_L(z,\bar z)],\autoeq$$
\newcount\localprod\localprod=\equno
\newcount\localprodS\localprodS=\secno
where $\nabla_i$ is the covariant derivative.

\bigskip
\equno=0
\secno=6
\centerline{\bf\S \autosec Fuzzy Derivatives}
\bigskip

The star product defined here can be used for more than just
multiplying functions on the fuzzy $\CPNF$, it can also be used
to define derivatives on the discrete fuzzy spaces.
In the continuum the vector fields on $\CPN$  generating $SU(N)$
can be expressed as
$${\cal L}_A=-i{f_{AB}}^C \xi^B\partial_C=i{1\over\sqrt{2}}{J_A}^C
\partial_C.\autoeq$$ 
\newcount\Liedefa\Liedefa=\equno
\newcount\LiedefaS\LiedefaS=\secno
It is easily verified that
$$[{\cal L}_A,{\cal L}_B]=i{f_{AB}}^C{\cal L}_C.\autoeq$$
\newcount\Lalg\Lalg=\equno
The corresponding action of a generator $L_A$ on the Hilbert space
${\cal H}_L$ is obtained from exponentiating the generator,
that is by considering $D_L(\eta)=e^{i\eta^AL_A}$:
$$Tr[\P_L(\xi)D_L(\eta^{-1})\hat F D_L(\eta)]
=Tr[\P_L(\xi_o)D_L(g^{-1}\eta^{-1})\hat F D_L(\eta g)].\autoeq$$
\newcount\Liedef\Liedef=\equno
Infinitesimally, with $\eta_A$ small and $D_L^{-1}(\eta)\approx 1-i\eta^AL_A$,
$$Tr[\P_L(\xi) (1-i\eta^AL_A)]=
\left\{Tr\left[\P_1(\xi)\left(1-i\eta^A\left({t_A\over\sqrt{2}}\right)\right)\right]\right\}^L\approx
1-{iL\over\sqrt{2}}\eta^A\xi_A,\autoeq$$
So $Tr[\P_L(\xi)L_A]={L\over\sqrt{2}}\xi_A$.
(The generators (\the\GMalgebraS.\the\GMalgebra) in the fundamental representation were normalised
so that $[{t_A\over\sqrt{2}},{t_B\over\sqrt{2}}]=i{f_{AB}}^C
{t_C\over\sqrt{2}}$.)

Now the derivative of a function in the continuum, ${\cal L}_AF(\xi)$,
can be taken over to the fuzzy case as
$$({\cal L}_A F_L)(\xi):=Tr\{\P_L(\xi)[L_A,\hat F]\}={L\over\sqrt{2}}(\xi_A*F_L-F_L*\xi_A).\autoeq$$
\newcount\Liedefb\Liedefb=\equno
\newcount\LiedefbS\LiedefbS=\secno
Using the $*$-product (6.12) this is
$$({\cal L}_A F_L)(\xi)={1\over\sqrt{2}}\bigl(K^{AB}\partial_BF_L-(\partial_BF_L)K^{BA}\bigr)
={i\over\sqrt{2}}J^{AB}\partial_BF_L,\autoeq$$
and this shows that the definition (\the\secno.\the\Liedefb) is consistent with (\the\secno.\the\Liedefa).
The main point here is that derivatives on functions in the continuum
restrict to derivatives at finite $L$ which can be represented as
commutators in the matrix algebra,
$$({\cal L}_A F_L) (\xi))=Tr\{\P_L(\xi)[L_A,\hat F]\}.\autoeq$$
\newcount\LieComm\LieComm=\equno
\newcount\LieCommS\LieCommS=\secno

This formula can now be used to prove that the symplectic form,
$\Omega$, defined in (\the\symplecticS.\the\symplectic) is
closed. Let ${\cal L}ie_X$ denote the Lie derivative along the vector
field $X$. Then, in the formula for the exterior derivative of a
2-form acting on three tangent vectors, $\X,\Y$ and $\Z$,
$$\eqalign{
d\Omega(\X,\Y,{\bf Z})=&{\cal L}ie_\X \Omega(\Y,{\bf Z}) + {\cal
L}ie_\Y \Omega({\bf Z},\X) +{\cal L}ie_{\bf Z} \Omega(\X,\Y) \cr
\qquad & -\Omega([\X,\Y],{\bf Z})-\Omega([\Y,{\bf Z}],\X) 
-\Omega([{\bf Z},\X],\Y),\cr}\autoeq$$
we represent all tangent vector fields by matrices as in
(\the\secno.\the\LieComm) (any tangent vector can be written as a
linear combination of the ${\cal L}_A$ at each $\xi$) and conclude
that $d\Omega=0$ by the Jacobi identity.

At this point, it is possible to derive simply the target space of the
mapping (6.2) from operators $\hat{F}$ on ${\cal H}_L$ to functions
$F_L(\xi)$ on $\CPN$.  Since the derivations $[\cdot,L_A]$ in ${\cal
H}_L$ are sent exactly to the derivations ${\cal L}_A$ in $\CPN$ by
the mapping, the second order Casimir in the adjoint action in ${\cal
H}_L$ is mapped to the Laplacian in $\CPN$, and the commutator actions
of the Cartan sub-algebra operators are sent to their equivalent
derivations in $\CPN$. This means that the normalised simultaneous
eigenvectors of all these Cartan operators in ${\cal H}_L$ are mapped
to simultaneous eigenfunctions of all the corresponding derivation
operators in $\CPN$ with the same eigenvalues.  Denoting the
irreducible tensor operators which are eigenvectors of the Cartan
operators by $\hat T^{\bf J}_{\bf M}$, with ${\bf J}$ a multiple index
labelling the representation and ${\bf M}$ a multi-index labelling the
weights, we find that $\hat T^{\bf J}_{\bf M}$ are mapped to $c^{\bf
J}(L)Y^{\bf J}_{\bf M}$, $Y^{\bf J}_{\bf M}$ being spherical
harmonics, the analogues of $Y^l_M$ for $SU(2)$. The constants $c^{\bf
J}(L)$ can easily be calculated and are all non-zero, which implies
the injectivity of the mapping $F_L$ assumed in (6.2). Thus, the
target of the mapping is just the space generated by the
eigenfunctions $Y^{\bf J}_{\bf M}$ of the Laplacian which are images
of the $\hat T^{\bf J}_{\bf M}$, with ${\bf J}$ running over all
$SU(N)$ irreducible representations in the $d_L\times d_L$ reducible
representation that contain $U(N)$ singlets.  For example ${\bf
CP}^1\cong S^2\cong SU(2)/U(1)$ requires $L$-fold symmetric
representations with $d_L=(L+1)$ and the $(L+1)\times (L+1)$ reducible
representation decomposes into irreducible representations as ${\bf
1}+{\bf 3}+\cdots +({\bf 2L+1})$.  There is only one Casimir for
$SU(2)$, so ${\bf J}$ is just the integer $l$ of the associated
irreducible representation and $M$ is the magnetic quantum number.
The $\hat T^l_M$, $l=0,\ldots,L$, are a basis for all
$(L+1)\times (L+1)$ matrices and $Y^l_M$ are the usual spherical
harmonics.

\bigskip
\equno=0
\centerline{\bf\S \autosec Conclusions}
\bigskip
The central result of this paper is equation (6.12),
which gives the explicit construction of an associative $*$-product
on the fuzzy $\CPNF$ between two functions \hbox{$F_L=Tr\{\P_L \hat
F\}$} and \hbox{$G_L=Tr\{\P_L \hat G\}$},
$$\eqalignno{
(F_L*&G_L)(\xi)=F_L(\xi)G_L(\xi)\cr
&+\sum_{l=1}^L{(L-l)!\over L!l!}
[\partial_{A_1}\cdots\partial_{A_l}F_L(\xi)]K^{A_1B_1}\cdots 
K^{A_lB_l}
[\partial_{B_1}\cdots\partial_{B_l}G_L(\xi)],}
$$
This expression is written in terms of an over-complete set of
coordinates $\xi^A$ in ${\bf R}^{N^2-1}$, with constraints
(\the\tlS.\the\tl). The projector $K=(P+iJ)/2$ in equation
(\the\KdefS.\the\Kdef) is defined by
$$P_{AB}={2\over N}\delta_{AB}+{\sqrt{2}}({d_{AB}}^C\xi_C)-2\xi_A\xi_B$$
and 
$$J_{AB}=\sqrt{2}{f_{AB}}^C\xi_C.$$
$P=-J^2$ is itself a projector mapping ${\bf R}^{N^2-1}$ onto the tangent
plane of $\CPN$ at $\xi^A$.

$P$ and $J$ are essentially the components of the usual hermitian
structure on $\CPN$ obtained by embedding it in the space of hermitian
matrices ${\bf R}^{N^2}$.  The latter is encapsulated in the three
equations (\the\complexS.\the\complex), (\the\metricS.\the\metric) and
(\the\symplecticS.\the\symplectic):
$$J(\M):=i[\P,\M],$$
$${\cal G}(\M_1,\M_2):=\Tr(-J^2(\M_1),\M_2)$$
and
$$\Omega(\M_1,\M_2):=\Tr(\M_1 J(\M_2))=-i\Tr(\P[\M_1,\M_2]),$$
describing the complex structure, the Fubini-Study metric and the
symplectic structure on $\CPN$ respectively. In our normalisation
convention $P=\G$.  Expressed in local holomorphic coordinates $z^i$,
$i=1,\ldots,N-1$, rather than the global coordinates, $\xi^A$, this
$*$-product is (\the\localprodS.\the\localprod),
$$\eqalignno{
(F_L*&G_L)(z,\bar{z})=F_L(z,\bar{z})G_L(z,\bar{z})\cr
&+\sum_{l=1}^L{(L-l)!\over L!l!} [\nabla_{{\bar j}_1}\cdots
\nabla_{{\bar j}_l}F_L(z,\bar z)](i\Omega^{{\bar j}_1 i_1}) \cdots 
(i\Omega^{{\bar j}_l i_l}) [\nabla_{i_1}\cdots\nabla_{i_l}
G_L(z,\bar z)].}$$
The $*$-star product reduces to the ordinary commutative product on
the continuous $\CPN$ in the $L\rightarrow\infty$ limit for fixed
$F_L$ and $G_L$ [\the\rieffel].

Note also the important expression for
the derivative of a function on the fuzzy $\CPNF$ as a 
commutator (\the\LieCommS.\the\LieComm),
which appears naturally in this construction
$$ {\cal L}_A F_L=Tr\{\P_L [L_A,\hat F]\}.$$

\bigskip
\centerline{\bf Acknowledgments}
\bigskip

It is a pleasure to thank Oliver Jahn for discussions, and in
particular for
pointing out the identity (\the\oiS.\the\oi) to us, and Peter
Pre\v{s}najder for useful comments.  We have benefited greatly from
participating in the study groups at CINVESTAV, Mexico D.F. and
Syracuse during the Summer and Fall of 2000, and the Summer of 2001.
This work was supported by DOE contract DE-FG02-85ERR40231, NSF grant
INT-9908763, joint NSF-CONACyT grant E120.0462/2000, CONACyT grant
30422-E and Enterprise Ireland, Basic Research grant
SC/1998/739. B.D. would also like to thank the Dublin Institute for
Advanced Studies for financial support.

\bigskip
\equnoA=0
\centerline{\bf Appendix A}
\bigskip

\Kprod=12
In this appendix we derive some essential properties of the matrix
$K=(K_{AB})$ used in the definition of the $*$-product
(\the\KprodS.\the\Kprod).  First we show that $K$ is a projector, with
rank $N-1$. To this end break $K$ into real and imaginary parts as in
the text, $K={1\over 2}(P+iJ)$ with
\equnoA=0
$$P_{AB}:={2\over N}\delta_{AB}-2\xi_A\xi_B+\sqrt{2}{\cal S}_{AB}
\autoeqA$$\newcount\Pdef\Pdef=\equnoA
and 
$$J_{AB}:=\sqrt{2}{\cal A}_{AB},\autoeqA$$\newcount\Jdef\Jdef=\equnoA
with symmetric matrix ${\cal S}_{AB}:={d_{AB}}^C\xi_C$ and 
the anti-symmetric matrix ${\cal A}_{AB}:={f_{AB}}^C\xi_C$
(all indices are raised and lowered here using $\delta_{AB}$).
It is shown in the text that $J$ corresponds to the complex 
structure on
${\bf CP}^{N-1}$, and we show here that $-J^2$ is
a projector of rank $2(N-1)$, with
$P J=J P=J$ and finally $J^2=-P$, which implies in particular that $P$
itself is also a projector. 

\noindent i) {\it $K$ is a projector with rank $N-1$}. To see this 
observe that 
$$Tr(t_At_Bt_Ct_D)={1\over N}\delta_{AB}\delta_{CD}
+{1\over 2}\bigl({d_{AB}}^E+i{f_{AB}}^E\bigr)
\bigl(d_{ECD}+if_{ECD}\bigr).\autoeqA$$
Now contracting this with $\xi^C\xi^D$ and using cyclic symmetry of the trace
and the constraints (\the\tlS.\the\tl) yields the two identities: 
$${\cal S}^2_{AB}-{\cal A}^2_{AB}=
{2(N-1)\over N^2}\delta_{AB}-{2\over N}\xi^A\xi^B
+{\sqrt{2}(N-2)\over N}{\cal S}_{AB}\autoeqA$$
\newcount\fourtrace\fourtrace=\equno
and
$$({\cal S}{\cal A}+{\cal A}{\cal S})_{AB}={\sqrt{2}(N-2)\over N}{\cal A}_{AB}.\autoeqA$$
\newcount\fourim\fourim=\equno
From these
it follows easily that $K^2=K$. Since the constraints also
dictate that $tr(K)=N-1$ ($tr$ here
means trace over the adjoint representation of $SU(N)$, so 
${\delta_A}^A=N^2-1$), $K$ is a projector onto an $N-1$ dimensional subspace of ${\bf R}^{N^2-1}$. 
\bigskip
\noindent ii) {\it $J^2$ is a projector and $J^3=-J$}. 
In the text the complex structure was denoted by $J$, and we can identify
that with the symplectic structure when the normalisation is such that
indices are raised and lowered with $\delta_{AB}$.
For completeness we give here an alternative derivation. 
First we show that 
$J^3=-J$ and $tr(-J^2)=2(N-1)$. 
By definition $J_{AB}:=\sqrt{2}{\cal A}_{AB}=\sqrt{2}f_{ABC}\xi^C$,
so $J_{AB}t^B=i[t_A,\xi]$ where ${\bf\xi}=\xi^At_A$. The constraints
(\the\tlS.\the\tl) imply 
$$\xi^2=\left({N-1\over N^2}\right){\bf 1}
+\left({N-2\over N}\right)\xi.\autoeqA$$
\newcount\xisquare\xisquare=\equnoA
Using the commutation
relations for $t_A$ gives
$$[[[t_A,\xi],\xi],\xi]=i(J^3)_{AB}t^B,\autoeqA$$
while expanding the commutators on the left hand side explicitly,
and using (A.\the\xisquare),
gives
$$[[[t_A,\xi],\xi],\xi]=-iJ_{AB}t^B,
\autoeqA$$
from which we conclude that $J^3=-J$. This means that $-J^2$ is
a projector since $(-J^2)^2=(-J^2)$ and the definition of $J$, (A.\the\Jdef),
together with the constraints (\the\tlS.\the\tl) and the
standard normalisation $f_{ACD}f_{BCD}=N\delta_{AB}$, show that
$tr(-J^2)=2(N-1)=dim\CPN$.
\bigskip
\noindent iii) {\sl $J$ commutes with $P $ and $P J=J$}. 
Since $d_{ABC}$ is an invariant tensor we have
$${f_{AB}}^Hd_{HCD}+{f_{AC}}^Hd_{BHD}+{f_{AD}}^Hd_{BCH}=0$$
and contracting this with $\xi^A\xi^B$ shows that
${\cal S}$ commutes with ${\cal A}$, since $f_{ABC}$ is totally
antisymmetric. The latter also means that $J$ annihilates $\xi$, so
that $J$ commutes with $P$. Since $K^2=K$ we have $J=(PJ+J P)/2$,
and hence $PJ=J$. 
\bigskip
\noindent iv) {\it $P=-J^2$}. The real part of $K^2=K$ implies
that $P^2-J^2=2P$. Since $P$ commutes with $J$ they
are simultaneously diagonalisable and because $-J^2$ is a projector, 
its eigenvalues are all $0$ or $1$. So the eigenvalues of $P$ are
$1$ when the eigenvalue of $-J^2$ is $1$, and either $0$ or $2$ when
the eigenvalue of $-J^2$ is $0$. Calling $p$ the number of eigenvalues
$2$ in $P$,we have
$$tr(P)=tr(-J^2)+2p,\autoeqA$$
while, directly from the definition of $P$ (A.\the\Pdef)
and the constraint equations (4.\the\tl),
one finds 
$$tr(P)=2(N-1)=tr(-J^2)\autoeqA$$
which implies that $p=0$. Thus we have $P=-J^2$, with
$P$ a projector of rank $2(N-1)$. Note that this implies
that $K$ annihilates the coordinates $K_{AB}\xi^B=0$, since $J$ does,
which is easily checked using (A.\the\Jdef).

\bigskip
\equnoB=0
\centerline{\bf Appendix B}
\bigskip

\genKdefS=5
In this appendix we give an alternative, more detailed, 
proof of the identity (\the\oiS.\the\oi),
$$K^{AB}K^{CD}\partial_B K_{DE}=0.\eqno (\the\oiS.\the\oi)$$
Denoting the generators of $SU(N)$ in the adjoint representation by $(\theta_A)_{BC}=-if_{ABC}$,
with commutation relations
$[\theta_A,\theta_B]=if_{ABC}\theta_C$, we have $J=i\sqrt{2}\theta_A\xi^A$ and
$$J_{AB}\partial_B J=\sqrt{2}i[\theta_A,J],\qquad\qquad
J_{AB}\theta_B=[\theta_A,J].\autoeqB$$
Using these commutators it is straightforward to show that
$$ K_{AB}\partial_B K = {1\over\sqrt{2}}({\bf 1}+iJ)_{AB}[K,\theta_B]=
{1\over\sqrt{2}}\left( [K,[\theta_A,iJ]]+[K,\theta_A]\right).\autoeqB$$
Now, since $K^2=K$ we have $K(dK)+(dK)K=dK$ from which 
$$K(dK)=dK(1-K).\autoeqB$$
The eigenvalues of $iJ$ are $\pm 1$ (each with multiplicity $(N-1)$) and
$0$ (with multiplicity $(N-1)^2$). We can thus choose a basis where 
$$iJ=\left(\matrix{{\bf 1}_{(N-1)} &&\cr &{\bf 0}_{(N-1)^2}&\cr&&-{\bf 1}_{(N-1)}\cr}\right)
\autoeqB$$
and
$$K=\left(\matrix{{\bf 1}_{(N-1)} &&\cr &{\bf 0}_{(N-1)^2}&\cr&&{\bf 0}_{(N-1)}\cr}\right)
=\left(\matrix{{\bf 1}_{(N-1)} &&\cr &&{\bf 0}_{N(N-1)}\cr}\right)
\autoeqB$$
where, for example,
${\bf 1}_{(N-1)}$ is the $(N-1)\times (N-1)$ identity matrix and ${\bf 0}_{(N-1)}$
the $(N-1)\times (N-1)$ square matrix with all entries zeros.  In terms of the $2\times 2$
block structure of the second form of $K$ above, we write 
$dK=\left(\matrix{{\bf A}&{\bf B}\cr {\bf C}&{\bf D}\cr}\right)$.  Equation (B.3) then shows that
$$KdK=\left(\matrix{{\bf 0}&{\bf B}\cr {\bf 0}&{\bf 0}\cr}\right),$$
so we only need examine
$<1|KdK|0>$ and $<1|KdK|-1>$ where $iJ|n>=n|n>$.
Now from (B.2) 
$$ K K_{AB}\partial_B K =
{1\over\sqrt{2}}K \left( [[iJ,\theta_A],K]+[K,\theta_A]\right),\autoeqB$$
and, since $K|1>=|1>$, $K|0>=K|-1>=0$, we deduce that
$$
<1|K K_{AB}\partial_B K|0>={1\over\sqrt{2}}<1| [\theta_A,iJ]+\theta_A|0>=0
\autoeqB$$
$$
<1|K K_{AB}\partial_B K|-1>={1\over\sqrt{2}}<1|([\theta_A,iJ] +\theta_A)|-1>
=-{1\over\sqrt{2}}<1|\theta_A |-1>.\autoeqB
$$
The last expression vanishes, because $\theta_A$ does not connect $|1>$ and $|-1>$,
and (\the\oiS.\the\oi) follows.

\bigskip

\centerline{\bf References}
\bigskip
\item{[\the\Connes]} A.~Connes, {\sl  Non-commutative Geometry}, Academic Press, (1994)

\item{[\the\Madore]} J.~Madore, {\sl An Introduction to Non-commutative Differential Geometry
and its Applications},
 CUP, (1995), {\tt gr-qc/9906059}; \hfill\break
G.~Landi, {\it An Introduction to Non-commutative Spaces And Their Geometries},
Springer-Verlag, Berlin, 1997, {\tt hep-th/9701078};\hfill\break
J.M.~Garcia-Bondi\'a, J.C.~Varilly and H.~Figueroa, {\sl Elements of non-commutative
Geometry}, Birkh\"auser (2001).

\item{[\the\European]} H. Grosse, C. Klim\v{c}\'{\i}k and  P. Pre\v{s}najder, {\it Int. J. Theor. Phys.} {\bf 35} 231-244, (1996) {\tt hep-th/9505175}

\item{[\the\Presnajder]} P.~Pre\v{s}najder, {\sl The Origin of Chiral Anomaly and the Non-commutative Geometry}, hep-th/9912050; the supersymmetric
generalisation of the $*$-product in this paper has been done by 
A.P.~Balachandran, Seckin Kurkcuoglu and Efrain Rojas Marcial (in preparation).   

\item{[\the\gropre]} H.~Grosse and P.~Pre\v{s}najder, {\it Lett. Math. Phys.} {\bf 33}, 171 (1995)
and references therein.

\item{[\the\grklpra]}
H.~Grosse, C.~Klim\v{c}\'{\i}k and P.~Pre\v{s}najder,
{\it Comm. Math. Phys.} {\bf 178},507 (1996); {\bf
185}, 155 (1997);
H.~Grosse and P.~Pre\v{s}najder, 
{\it Lett. Math. Phys.} {\bf 46}, 61 (1998) and ESI preprint,
1999.  

\item{[\the\grklprb]}
H.~Grosse, C.~Klim\v{c}\'{\i}k, and P.~Pre\v{s}najder,
{\tt hep-th/9602115} and
{\it Comm. Math. Phys.} {\bf 180}, 429 (1996).

\item{[\the\grklprc]}
H.~Grosse, C.~Klim\v{c}\'{\i}k, and P.~Pre\v{s}najder,
in {\it Les Houches Summer School 
on Theoretical Physics}, 1995,
{\tt hep-th/9603071}.

\item{[\the\pr]}
P.~Pre\v{s}najder,
{\tt hep-th/9912050} and 
{\it J. Math. Phys.} {\bf 41} (2000) 2789-2804.

\item{[\the\watamuraa]}
U.~Carow-Watamura and S.~Watamura,
{\tt hep-th/9605003} and 
{\it Comm. Math. Phys.} {\bf 183}, 365 (1997).
 \item{[\the\watamurab]}
U.~Carow-Watamura and S.~Watamura,
{\tt hep-th/9801195} and {\it Comm. Math. Phys.} {\bf 212}, 395 (2000).

\item{[\the\bbiv]}
S.~Baez, A.~P. Balachandran, S.~Vaidya and B.~Ydri,
{\tt hep-th/9811169} and {\it Comm. Math. Phys.} {\bf 208}, 787 (2000). 

\item{[\the\Bala]}
A.P.~Balachandran and S.~Vaidya,
{\tt hep-th/9910129} and 
{\it Int. J. Mod. Phys.} {\bf A16}, 17 (2001);
A.~P.~Balachandran, T.~R.~Govindarajan and B.~Ydri,
{\tt hep-th/9911087};
A.~P.~Balachandran, T.~R.~Govindarajan and B.~Ydri, 
{\tt hep-th/0006216} and {\it Mod. Phys. Lett.} 
{\bf A15}, 1279 (2000);
A.~P.~Balachandran, X.~Martin and D.~O'Connor, {\tt hep-th/0007030}
and {\it Int. J. Mod. Phys.} {\bf A} (in press);
S.~Vaidya, {\tt hep-th/0102212}; Chong-Sun~Chu, J.~Madore and H.~Steinacker, {\tt hep-th/0106205}.

\item{[\the\grost]} 
Fuzzy ${ C}{ P}^{2}$ has already been
investigated by H.Grosse and A.Strohmaier,\hfill\break
{\it hep-th/9902138}
and {\it Lett. Math. Phys.} {\bf 48}, 163 (1999).

\item{[\the\Balb]} G. Alexanian, A.P. Balachandran, G.Immirzi and B.Ydri, {\sl Fuzzy $CP^2$}, {\it J. of Geom. and Phys.} (in press) , {\tt hep-th/0103023}

\item{[\the\bdmo]}
G.~Alexanian, S.~Pinzul and A.~Stern {\tt hep-th/0010187} and
{\it Nucl. Phys.} {\bf B600/3}, 531 (2001)
\item{[\the\UG]} Brian Dolan and Oliver Jahn, in preparation.

\item{[\the\strings]} C.~Klim\v{c}\'{\i}k and P.~\v{S}evera, {\tt hep-th/9609112} and
{\it Nucl. Phys.} {\bf B383} (1996) 281; \hfill\break
A.Y.Alekseev and V.Schomerus, 
{\tt hep-th/981112193} and H.~Garcia-Compean and J.F.~Plebanski, 
{\tt hep-th/9907183} and K.~Gawedzki, {\tt hep-th/9904145};\hfill\break
R.C.~Myers, {\tt hep-th/9910053} and JHEP 9912,022 (1999);
S.P.~Trivedi and S.~Vaidya, {\tt hep-th/0007011} and 
JHEP {\bf 0009}, 41 (2000);
S.R.~Das, S.P.~Trivedi and S.~Vaidya, {\tt hep-th/0008203}
and JHEP {\bf 0010}, 037 (2000);\hfill\break
S.~Ramgoolam, {\sl On spherical harmonics for fuzzy spheres in diverse dimensions}, 
{\tt hep-th/0105006};
Z.~Guralnik, S.~Ramgoolam, JHEP {\bf 0102}, 032 (2001), {\tt hep-th/0101001}

\item{[\the\Kontsevich]} M.~Kontsevich, {\sl Deformation quantisation of Poisson manifolds}, 
{\tt q-alg/9709040};\hfill\break
A.S.~Cataneo and G.~Felder {\it Mod. Phys. Lett.} {\bf A16}, 179 (2001), {\tt hep-th/0102208};
A.S.~Cataneo and G.~Felder {\it Comm. Math. Phys.} {\bf 212}, 591 (2000), {\tt math.QA/9902090}

\item{[\the\rieffel]} M.A. Rieffel, {\tt math.QA/0108005}. 

\item{[\the\ksp]}
J.R.~Klauder and B.-S.~Skagerstam, {\it Coherent States: Applications in
Physics and Mathematical Physics}, World Scientific (1985);
A.M.~Perelomov: {\it Generalized Coherent States and their Applications},
Springer-Verlag, (1986);
M.~Bordemann, M.~Brischle, C.~Emmrich and  S.~Waldmann,
{\it J. Math. Phys.} {\bf 37}, 6311 (1996);
M.~Bordemann, M.~Brischle, C.~Emmrich and  S.~Waldmann,
{\it Lett. Math. Phys.} {\bf 36}, 357 (1996); 
S.~Waldmann, {\it Lett. Math. Phys.} {\bf 44}, 331 (1998).

\item{[\the\Wedderburn]} See {\it e.g.} theorem 6.3.8 in G.~Murphy, 
{\sl $C^*$ Algebras and Operator Theory.} Academic Press (1990)

\item{[\the\Mukunda]} N.~Mukunda, Arvind, S.~Chaturvedi and R.~Simon, 
{\sl Generalised coherent states and the diagonal representation for operators.}
 {\it Int. J. Mod. Phys.} {\bf A} (in press),\hfill\break
\hbox{\tt quant-ph/0002070}

\bye